\documentclass[twocolumn,aps,prl,superscriptaddress]{revtex4}
\usepackage{graphicx}
\usepackage{amsfonts}
\usepackage[varg]{txfonts}
\usepackage{bm}

\begin{document}

\title{Topological Insulators and Metals in Atomic Optical Lattices}
\author{Tudor D. Stanescu}
\affiliation{Joint Quantum Institute and Condensed Matter Theory Center, 
Department of Physics, University of Maryland, College Park, MD 20742-4111}
\author{Victor Galitski}
\affiliation{Joint Quantum Institute and Condensed Matter Theory Center, 
Department of Physics, University of Maryland, College Park, MD 20742-4111}
\author{J.Y. Vaishnav}
\affiliation{Joint Quantum Institute, National Institute of Standards and Technology, Gaithersburg MD 20899 USA}
\author{Charles W. Clark}
\affiliation{Joint Quantum Institute, National Institute of Standards and Technology, Gaithersburg MD 20899 USA}
\author{S. Das Sarma}
\affiliation{Joint Quantum Institute and Condensed Matter Theory Center, 
Department of Physics, University of Maryland, College Park, MD 20742-4111}

\begin{abstract}

We propose the realization of topological quantum states with cold atoms trapped in an optical lattice. We discuss an experimental setup that generates a two-dimensional  hexagonal lattice in the presence of a light-induced periodic vector potential, which represents a realization of the Haldane model with cold atoms.   We determine theoretically the conditions necessary for observing the topological states and show that two of the key conditions are: 1) the realization of sharp boundaries and 2) the minimization of any smoothly varying component of the confining potential. We argue that, unlike their condensed matter counterparts, cold atom topological quantum states can be i) ``seen'', by mapping out the characteristic chiral edge states, and ii) controlled, by controlling the periodic vector potential and the properties of the confining potential.

\end{abstract}


\maketitle

A solid state insulator can be defined as a system with local electronic properties~\cite{Kohn}. Consequently, insulators are insensitive to boundary conditions. Standard band insulators, which are characterized by the existence of a bulk energy gap, satisfy this definition and are topologically equivalent, as they can be adiabatically transformed into each other without crossing a phase transition. However, the existence of a bulk gap is not a sufficient condition to insure the locality of all electronic properties and the insensitivity to boundary conditions. Certain strongly correlated systems, such as the fractional quantum Hall fluids, offer examples of phases having bulk gaps but being topologically distinct~\cite{Wen}. Such topological insulators can exist even in the absence of interactions and typical examples are the integer quantum Hall fluids or the quantum spin Hall states~\cite{Thouless,Haldane,KaneMele} and their three dimensional generalizations~\cite{FuKane}. One defining characteristic of these systems is the existence of chiral gapless edge states that are robust against disorder effects and interactions. The existence of these edge states and their basic features are intrinsically linked to the topological properties of the system. However, their detailed structure is dictated by the boundary. Controlling the boundary properties is a rather difficult task for solid state systems, but it could be realized for ultracold atoms in optical lattices. 

In this work we propose the realization of a topological insulator in an optical lattice. More importantly, we establish that optical lattices allow for the existence of a ``topological metal'' characterized by a chiral boundary-induced edge mode, although the bulk is metallic. A great advantage of the optical lattices is that various terms in the Hamiltonian as well as the boundary conditions can be explicitly controlled experimentally, which in contrast to solid state systems, allows for tuning the properties of the edge states.
The realization of a topological insulator with cold atoms opens a series of very exciting prospects: i) the possibility of a direct observation of the edge states~\cite{Scarol}, the hallmark of the topological insulating phase, ii) the possibility of testing the stability of the chiral edge modes in the presence weak disorder and interactions, and iii) the possibility of studying  transitions between a topological insulator and other phases.  On the other hand, the  main challenges in building a topological insulator with cold atoms are: i) generating the vector potential, ii) controlling the trap potential and manipulating the boundaries, and iii) measuring a topological insulating state, i.e., imaging the edge states.

To study the boundary effects in a topological quantum state, we propose the realization of the Haldane model~\cite{Haldane} in an optical lattice. The trapping potential is given by the superposition of three co-planar standing waves characterized by the wave-vectors ${\bf k}_1 = (0, \frac{2\pi}{3a})$,  ${\bf k}_2 = (\frac{\pi}{\sqrt{3}a}, \frac{\pi}{3a})$  and ${\bf k}_3 = (-\frac{\pi}{\sqrt{3}a}, \frac{\pi}{3a})$, respectively. The minima of this potential generate a hexagonal lattice with lattice constant a. The crucial ingredient of the setup is an effective vector potential $\bf{A}({\bf r})$ that generates a periodic ``magnetic'' field with zero total flux trough the unit cell. We propose the use of a light-induced gauge potential that can be realized in a system of multi-level atoms interacting with a spatially modulated laser field~
\cite{Dum,Dutta,Jaksch,oberg1,Rusec,Osterloh,Soren,Zhu,Clark,SZG}.
Within these schemes, the multi-level atoms interact with laser beams characterized by spatially varying Rabi frequencies and experience an effective pseudo-spin dependent gauge potential. Since our proposal does not require spin-dependent gauge potentials, one may even be able to utilize a simpler scheme~\cite{Spiel}.
The effective single particle Hamiltonian is
\begin{equation}
H = \frac{1}{2m}\left({\bf p} - {\bf A}({\bf r})\right)^2 + V_0\sum_{i=1}^3 \cos^2({\bf k}_i{\bf r}) + V_c({\bf r}), 
\label{Ham}
\end{equation}
where $m$ is the particle mass, $p$ the momentum and ${\bf A}$ the vector potential. The second term in Eq. (\ref{Ham}) generates the two-dimensional hexagonal optical lattice with lattice constant $a$, while the last term, $V_c$, contains additional confining terms that determine the properties of the  boundaries. The role of $V_c$ will be discussed in detail below. A simple vector potential that generates zero ``magnetic'' flux trough a unit cell is ${\bf A}({\bf r}) = \alpha {\cal A}({\bf r})$ with ${\cal A}_x({\bf r}) = \sin\left(\frac{4\pi y}{3}\right)$ and ${\cal A}_y = 0$, where $\alpha$ is the strength of the gauge potential. Notice however that the term ${\bf A}^2/2m$ from Eq. (\ref{Ham}) represents an extra contribution to the lattice potential that does not have hexagonal symmetry and, therefore, will distort the lattice. This distortion does not affect the nature and the basic properties of the topological insulator if $\alpha$ does not exceed a certain critical value, but has quantitative implications for the band structure. To avoid the formal complications of dealing with this $\alpha$-dependent lattice distortion, we use in our calculations a symmetrized vector potential with 
${\cal A}({\bf r}) = \left[\sin(4\pi y/3) ~+~\cos(2\pi x/\sqrt{3})\sin(2\pi y/3),  ~~~-\sqrt{3}\sin(2\pi x/\sqrt{3})\right.$  $\left.\times\cos(2\pi y/3)\right]$. The fact that ${\bf A}^2/2m$ acts as an effective lattice potential suggests that ideally the trapping and vector potentials could be realized using the same set of laser fields.
\begin{figure}[tbp]
\begin{center}
\includegraphics[width=0.45\textwidth]{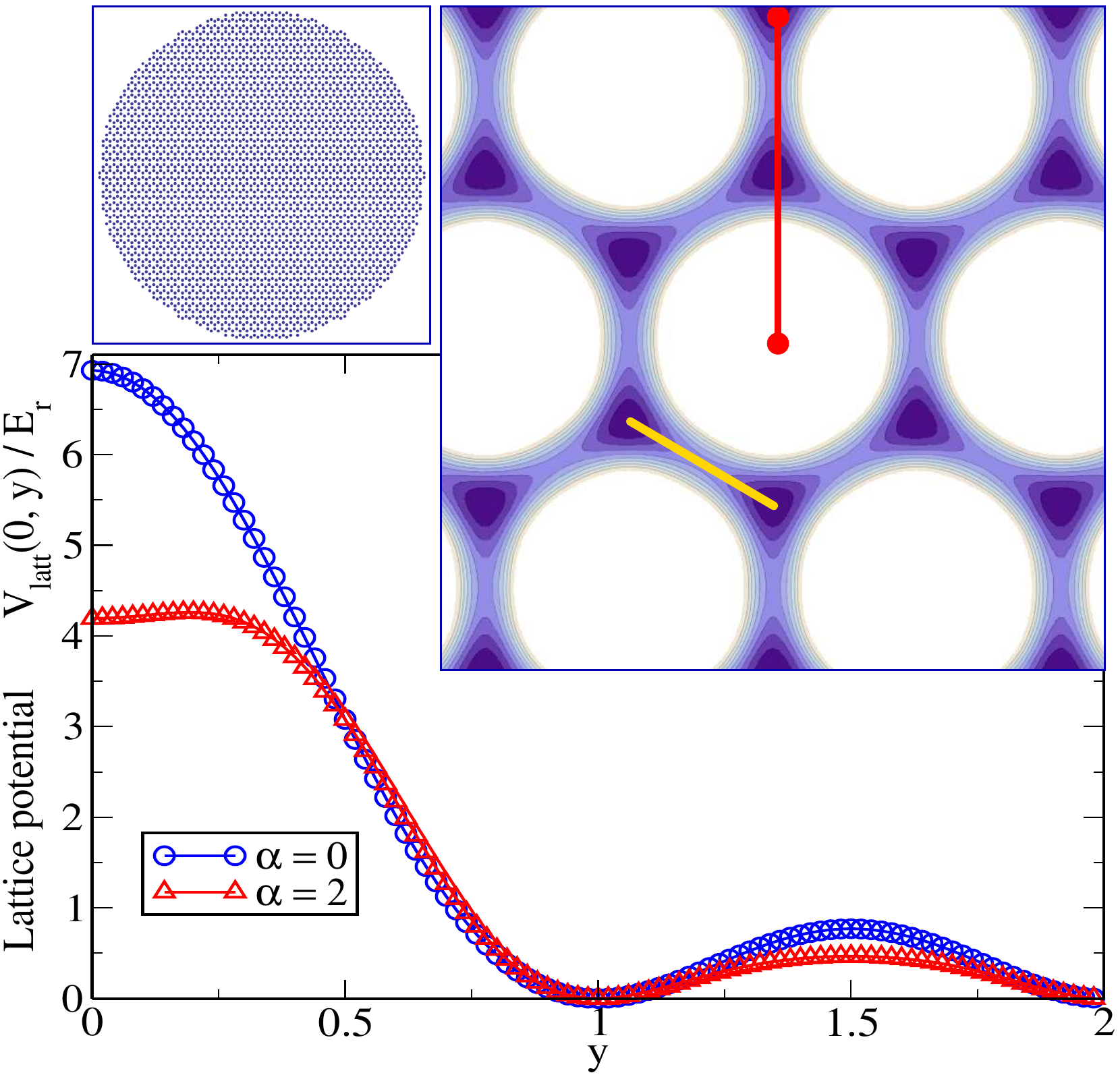}
\end{center}
\caption{Optical lattice potential formed by the superposition of three standing waves and generating a two-dimensional hexagonal lattice with lattice constant $a$. The effective confining potential along the segment $(0,0)-(0,2a)$ (red line in the upper right panel) is shown in the absence of a vector potential (blue circles) and for ${\bf A}\neq 0$ (red triangles). Inset: typical cluster used in the calculations consisting of a disc-shaped piece of hexagonal lattice with radius $R\approx 39a$. \vspace*{-0.15in}} \label{Fig1}
\end{figure}

We solve the quantum problem associated with Hamiltonian (\ref{Ham}) within a simple tight binding approximation. Throughout the paper we use the recoil energy $E_r = \frac{(\pi / a)^2}{2m}$ as energy unit and the lattice constant $a$ as length unit. The total effective lattice potential $V_{latt}$, which includes the term  ${\bf A}^2/2m$, with minima at the nodes of a hexagonal lattice (see Fig. \ref{Fig1}) has near these minima the form $V_{latt}\approx m\omega_0^2/2(\delta x^2 + \delta y^2)$. This  suggests the use of the s-wave orbitals $\phi_0^{(i)}({\bf r})= \sqrt{2/(\pi c)}\exp[-({\bf r}-{\bf r}_i)^2/c]$ as a possible simple basis for the tight-binding approximation. Here ${\bf r}_i$ represents the position of a lattice site, $c = (4 E_r)/(\pi^2 \omega_0) a^2$ and we have 
$V_0 = 12 E_r/\pi^4(a^4/c^2 - \pi^2\alpha^2a^2/4)$. The approximation holds as long as the s-band is well separated from the p-bands, which is the case for $c<0.25$, i.e., for deep enough optical lattices. However, because the the second-neighbor hopping is crucial for generating the topological states ~\cite{Haldane}, a small value of $c$ will make this effect practically unobservable. In this study we choose $c=0.2$. The other independent parameter is $\alpha$, which in the calculations will be either $\alpha=0$ (zero vector potential) or  $\alpha=2$. The hopping parameters for the effective tight-binding model are given by $t_{ij}= \langle\phi_0^{(i)}|H|\phi_0^{(j)}\rangle$. The key contributions coming from the vector potential, $\langle\phi_0^{(i)}|{\bf p}{\bf A}|\phi_0^{(j)}\rangle$, vanish if i and j are nearest neighbors and are  non-zero for second-order neighbors, generating a chiral contribution to the Hamiltonian.  
\begin{figure}[tbp]
\begin{center}
\includegraphics[width=0.48\textwidth]{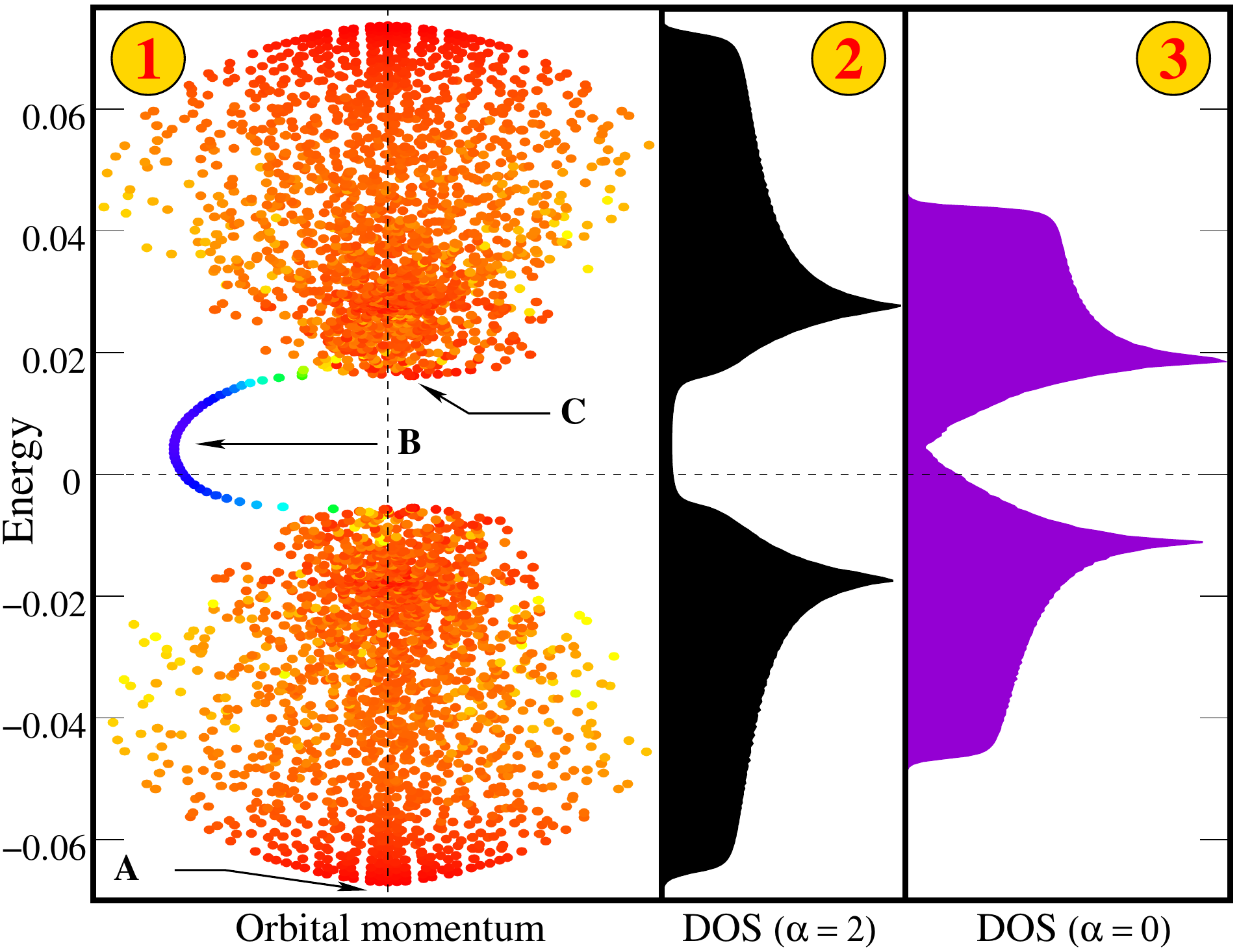}
\end{center}
\caption{Finite size equivalent of the Hofstadter ``moth'' for a cluster with a hard-wall boundary in the presence of a periodic vector potential with $\alpha=2$. The coordinates of each dot represent the orbital momentum (x-axis) and the energy (y-axis) of a particular state. The ``edge'' vs. ``bulk'' character of each state is shown by the color code, which represents the relative boundary contribution to the norm,  $\gamma_n$ (see main text). $\gamma_n$ ranges from 1 for edge states (blue) to 0 for purely bulk states (red). Notice the chiral nature of the edge states mode that populates the bulk gap of the insulator. The corresponding density of states (DOS) in shown in panel (2). For comparison, in panel (3) we show the DOS in the absence of a gauge potential. \vspace*{-0.15in}} \label{Fig2}
\end{figure}
Finally, because the s-wave orbitals are not orthogonal, we have to determine their overlap matrix elements $\langle\phi_0^{(i)}|\phi_0^{(j)}\rangle$. The resulting tight-binding problem is solved numerically for clusters containing up to $3696$ sites (see Fig \ref{Fig1}).

We start by solving the tight-binding problem for a finite-size cluster consisting of a disc with hard-wall boundary conditions, $V_c(r) = \infty$ if $r>R$ and $V_c(r) = 0$ if $r<R$. The density of states (DOS) for this system is shown in Fig. \ref{Fig2} (panels 2 and  3). In the absence of a vector potential ($\alpha=0$ - panel 3), this quantity  is similar to the graphene DOS. The characteristic  V-shaped gap associated in the infinite lattice with the existence of the Dirac points is weakly distorted at low energies by finite size effects that become negligible as the cluster size increases. In the presence of a vector potential with $\alpha = 2$ a finite size gap opens in the density of states (panel (2) Fig. \ref{Fig2}). 
\begin{figure}[tbp]
\begin{center}
\includegraphics[width=0.48\textwidth]{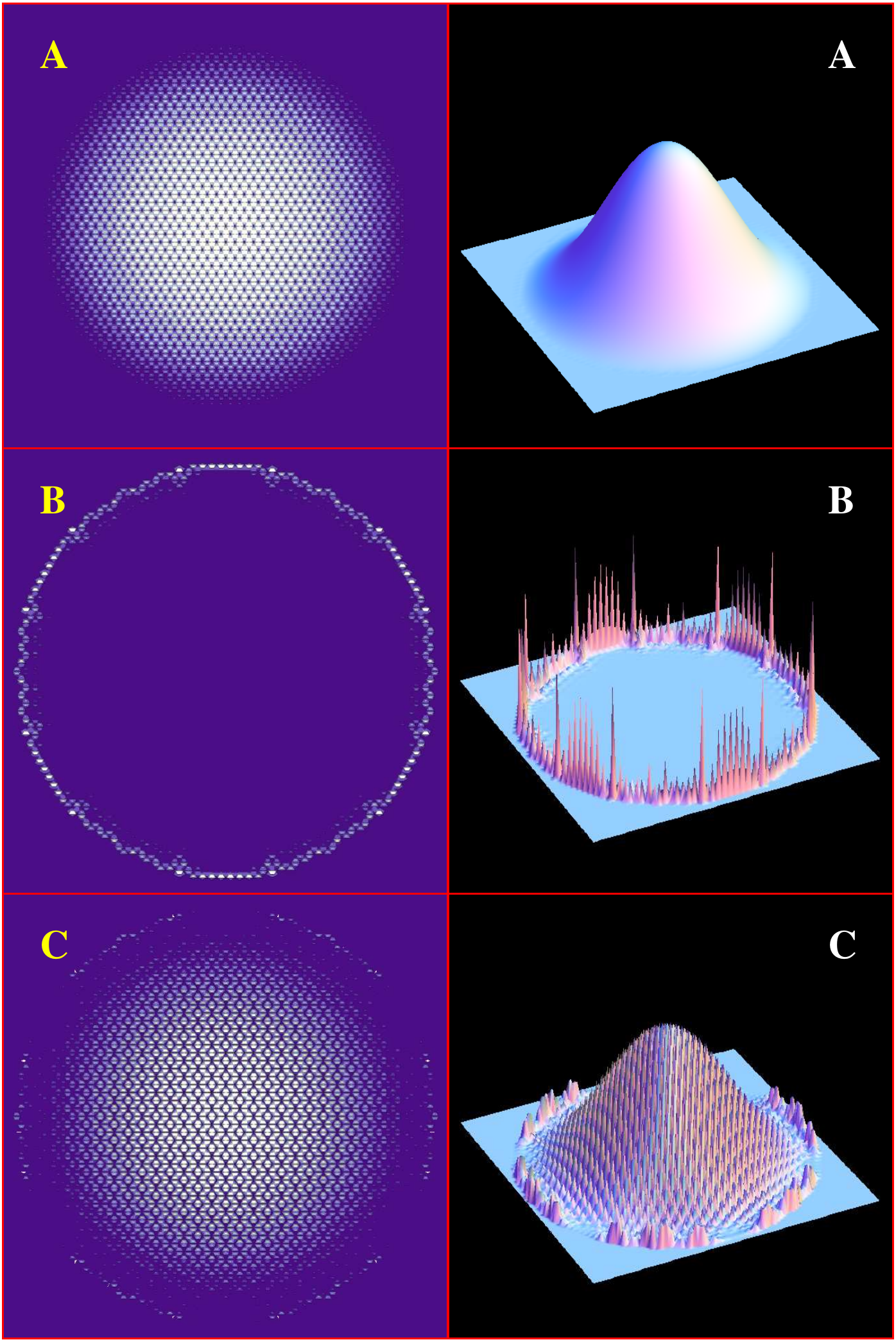}
\end{center}
\caption{Left panels: Contour plots of $\rho_n({\bf r})=\vert\psi_n({\bf r})\vert^2$ for the states marked in Fig. \ref{Fig3}. The quantity $\rho_n({\bf r})$ is the product between a common factor associated with the underlying hexagonal lattice structure and a state-dependent envelope function.  The corresponding envelope functions are shown in the right panels. (A) represents the ground state, (B) is a typical edge state and (C) is the lowest energy bulk state from the upper band. The weak edge contributions in (C) are due to finite size effects and vanish in the large cluster limit. \vspace*{-0.15in}} \label{Fig3}
\end{figure}
However the DOS in the ``gap'' is not exactly zero. To determine the nature of the residual in-gap states, we calculate the orbital momentum of each single-particle state $\psi_n({\bf r})$ and the relative contribution $\gamma_n$ to the norm $\langle\psi_n|\psi_n\rangle$ coming from a narrow ring $37a\leq r\leq 39a$ positioned at the edge of the system. This contribution vanishes for bulk-type states and is of order one for edge states, i.e., it represents a measure of the edge-type character of a given state. The corresponding spectrum is shown in Fig. \ref{Fig2} panel (1). This picture is the finite size equivalent of the  Hofstadter ``moth'' for a hexagonal lattice and a periodic ``magnetic'' field with zero flux through the unit cell. The coordinates of each dot represent the orbital momentum (x-axis) and the energy (y-axis) of a particular state. The edge vs. bulk character of the state, as quantified by the parameter $\gamma_n$ is revealed by the color code: blue for edge states and red for bulk states. As evident from Fig. \ref {Fig2}, the spectrum is characterized by a gap for the bulk (red) states. Within this gap, there is an chiral edge mode (blue states). The chirality of the edge mode, i.e., the sign of its  orbital momentum, is determined by the sign of $\alpha$. Note that the sum of the orbital momenta of all the single-particle states is identically zero for any value of $\alpha$.  
To have a spatial characterization of the single-particle quantum states, we show in Fig. \ref{Fig3} the contour plots of $\rho_n({\bf r})=\vert\psi_n({\bf r})\vert^2$ for several states marked in Fig. \ref{Fig2}: A - the ground state, B - a typical edge state and C  - the lowest energy bulk state from the upper band. Note that each density function $\rho_n({\bf r})$ consists of a common factor $\left[\sum_i\phi_0^2({\bf r}-{\bf r}_i)\right]^2$ associated with the underlying lattice structure  multiplied by a specific envelope function. The structure of the envelope functions is shown in the left panels of Fig. \ref{Fig3}.

\begin{figure}[tbp]
\begin{center}
\includegraphics[width=0.45\textwidth]{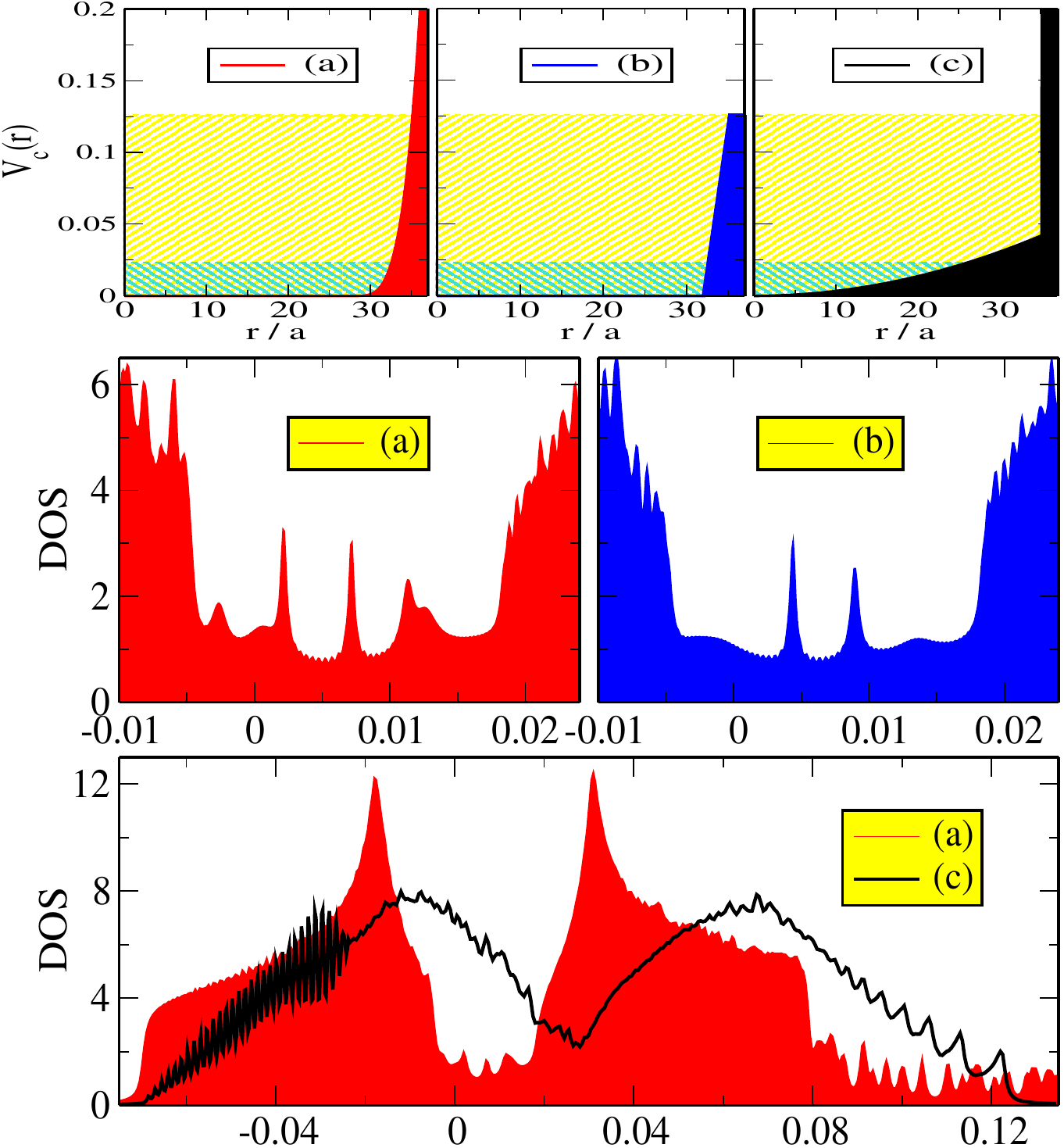}
\end{center}
\caption{System with soft boundaries. The confining potential is shown in the top panels for: a) quartic wall, b) linear step, and c) harmonic trap plus infinite wall. The turquoise region corresponds to energies smaller than the gap $\Delta_{\alpha}$, while in the yellow region the energy is smaller than the bandwidth $W$. A confining potential with characteristic length much smaller than the size of the system,(a) and (b), preserves the bulk gap (topological insulator). The in-gap features appearing in the density of states (middle panels) are all due to edge states (see also Fig. \ref{Fig5}). In a smoothly varying confining potential, case (c), the bulk gap collapses (topological metal, see lower panel). In cases (a) and (b) the lower band is unaffected by the details of the confining potential, while the in-gap structures (middle panels) are similar. The rapid oscillations at low-energies in case (c) indicate the formation of harmonic oscillator levels. \vspace*{-0.15in}} \label{Fig4}
\end{figure}
\begin{figure}[tbp]
\begin{center}
\includegraphics[width=0.45\textwidth]{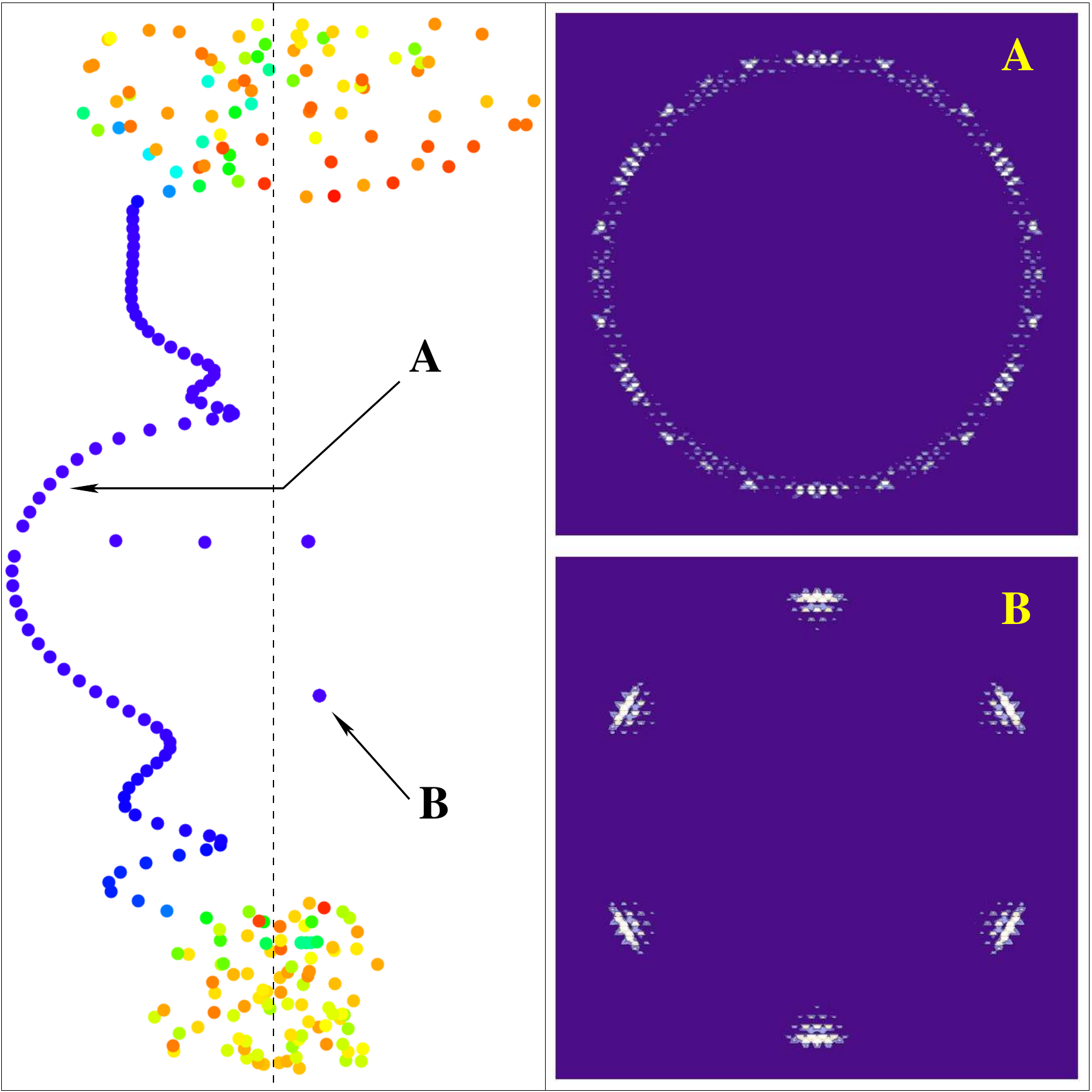}
\end{center}
\caption{In-gap states for a system with  soft boundary (quartic wall). The color code is the same as in figure Fig \ref{Fig2}. Note that all the in-gap states have edge -like character. In addition to the typical edge mode consisting of chiral  edge states (A), notice the presence of Tamm-like  edge states (B) with opposite orbital momentum. \vspace*{-0.15in}} \label{Fig5}
\end{figure}

Next, we address the key question concerning the role of the confining potential $V_c$ and the dependence of the spectrum on the boundary conditions. 
We replace the hard wall boundary by a) a quartic wall, b) a linear step potential, and c) a harmonic potential plus a hard wall (see Fig. \ref{Fig4}, top panel). The corresponding expressions of the confining potential are $V_c^a(r) = \lambda_c^a(r-R_0)^4$ if $r > R_0$ (and 0 otherwise), $V_c^b(r) = \mbox{Min}\left[\lambda_c^b(r-R_1)/(R_2-R_1), \lambda_c^b\right]$ if $r > R_1$ (and 0 otherwise), and $V_c^b(r) = \lambda_c^c r^2$ if $r < R_2$ (and $\infty$ otherwise), respectively. To define a characteristic length-scale associated with the confining potential, we introduce the radii $R_{\Delta}$, such that $V_c(R_{\Delta}) = \Delta_{\alpha}$ and $R_{W}$, with the property $V_c(R_{W}) = W$. Here  $\Delta_{\alpha}$ is the bulk gap and $W$  the sum of the bandwidths of the lower and upper bands for a system with hard walls. The relevant length scale for a soft boundary produced by the confining potential $V_c$ is given by $d_{c}=R_{W}- R_{\Delta}$. Our numerical calculations show that a topological insulator can be realized provided $d_{c}\ll R_{\Delta}$, i.e., the boundary has a characteristic length much smaller than the size of the system. For example, cases (a) and (b) in Fig. \ref{Fig5} correspond to $R_{\Delta}\approx 32.5 a$ and $d_{c}\approx 3.5a$ and in both cases the gap for bulk states is preserved. However, in contrast with hard wall case characterized by a featureless residual in-gap DOS (see Fig. \ref{Fig2}, panel 2), a system with soft boundaries has a nontrivial structure of the residual DOS (Fig. \ref{Fig4}, middle panels). This structure emerges from two causes: 1) the orbital momentum of the chiral edge mode acquires a more complicated energy dependence, and 2) additional edge states, that do not belong to the chiral edge mode, develop inside the gap. Both these points are illustrated in Fig. \ref{Fig5}. The B-type edge states are Tamm-like states, which are formed due to the rapid variation of the confining potential and are not related to the topological properties of the insulator. Finally, in the presence of a confining potential with a smoothly varying component (case (c) in Fig. \ref{Fig4}) the bulk gap collapses and the system becomes metallic. Note that, in the presence of the periodic vector potential ${\bf A}({\bf r})$, the system still has chiral edge states even in the metallic phase, a situation similar to the existence of surface states in doped semiconductors in the presence of spin-orbit coupling~\cite{SG}. We conclude that, in order to realize a topological insulator with cold atoms, one needs to produce a sharp boundary and to minimize any smoothly varying component of the confining potential.

Probing a topological state in condensed matter systems involves transport measurements. This would be a rather difficult task for atomic systems in optical lattices. Since the nontrivial topological properties of a system are a feature of the single particle Hamiltonian which is best revealed by the presence of the chiral edge states, we propose here the direct observation of these edge states in cold atomic systems, something one cannot easily realize in the condensed matter context. This involves two steps: 1) loading bosons into the edge states and 2) imaging the atoms.  Initially, the optical lattice is loaded with atoms and cooled, so that the bosons occupy only the lowest energy single particle states. Next, angular momentum is transferred to the atoms using, for example, a 2-photon stimulated Raman process~\cite{Andersen}, thereby populating states of higher energy and angular momentum comparable to that of the edge states. Finally, one can use resonant $\pi$ pulses to further excite the atoms into the edge states. To image the atoms, one can use a direct in-situ imaging technique~\cite{Nelson}. Although this technique is able, in principle, to discriminate between individual atoms in an optical lattice, a much lower resolution is required for imaging the edge states.

We propose the realization of topological quantum states with cold atoms in an optical lattice. A combination of a hexagonal optical lattice potential and a periodic light-induced vector potential represents a cold-atom realization of the Haldane model. We show that such a system is characterized by chiral edge modes, which are the signature of a topological quantum state. Observing these edge sates is an effective way of ``seeing'' a topological phase. We find that necessary conditions for realizing a topological insulator with optical lattices are the realization of a sharp boundary and the minimization of any smoothly varying component of the confining potential, e.g., of the harmonic confining potential. Controlling the confining potential opens the possibility of testing the stability of the chiral edge modes against weak disorder and interactions and, together with control of the vector potential, offers a knob for tuning the system from a topological insulator state to a standard insulator or a metallic phase.

This work is supported by NSF through JQI-PFC, DARPA and US-ARO.
\vspace*{-0.25in}

\bibliography{refsTopIns}

\end{document}